# THE GAME OF PHISHING

Joseph Kilcullen

Moylurg, Foxford Road, Ballina, Co. Mayo, F26 D9D2, Ireland.

*ABSTRACT*

*The current implementation of TLS involves your browser displaying a padlock, and a green bar, after successfully verifying the digital signature on the TLS certificate. Proposed is a solution where your browser's response to successful verification of a TLS certificate is to display a login window. That login window displays the identity credentials from the TLS certificate, to allow the user to authenticate Bob. It also displays a 'user-browser' shared secret i.e. a specific picture from your hard disk. This is not SiteKey, the image is shared between the computer user and their browser. It is never transmitted over the internet. Since sandboxed websites cannot access your hard disk this image cannot be counterfeited by phishing websites. Basically if you view the installed software component of your browser as an actor in the cryptography protocol, then the solution to phishing attacks is classic cryptography, as documented in any cryptography textbook.*

## 1. INTRODUCTION

Originally it was game theory research, seeking screening strategies, or signalling strategies, to prevent the counterfeiting of websites i.e. phishing attacks. Since your web browser is installed software, it is more capable e than the websites creating the counterfeit e.g. it can access the hard disk. Hence, various ways for websites to counterfeit installed software behaviour were studied. In full screen mode, it was found that, browsers can counterfeit almost anything, including blue screens of death and formatting the hard drive.

From an academic point of view, full screen counterfeiting eliminates several categories of installed software behaviour, as possible anti-counterfeiting solutions. One category of installed software behaviour was resistant to counterfeiting. Every solution, in that category, was found to be a user-browser shared secret. Basically Mallory cannot counterfeit what Mallory does not know. The user-browser shared secret is not known by either Bob or Mallory. Furthermore, such a simple solution prompted the following hypothesis. Web browsers are virtual machines. They execute each website inside a sandbox. Hence any given web browser has N + 1 personalities, at any given time. Where N is the number of webpages open i.e. one personality for each webpage, plus one for the installed software, of the browser itself. Once you view the installed software component of your browser as an actor in the cryptography protocol, the solution to phishing attacks becomes classic cryptography i.e. the installed software component, of your browser, must authenticate itself. It does this in the same way that cryptography actors have been authenticating themselves for thousands of years i.e. by presenting a previously shared secret. With that, game theory research was transformed into cryptography research.

## 2. CONCERNING THE CAPACITY OF BROWSERS TO COUNTERFEIT

**INSTALLED SOFTWARE BEHAVIOUR**

The idea is that a phishing attack is a game of incomplete information. That the user does not even know that a phishing attack is taking place. It is the successful counterfeiting of the website that does this. If we can devise a signalling strategy which cannot be counterfeited then the computer user will know when a phishing attack is taking place. They will back away from the phishing website causing the phishing attack to fail.

The idea was to add information, specifically an anti-counterfeiting signalling strategy which would be triggered after the browser has verified the digital signature on Bob's TLS certificate. I listed behaviour that installed software is capable of but websites are not capable of. The idea was: Your browser is installed software so it has this advantage over websites trying to counterfeit its behaviour. The following categories were proposed for research:

1. Drawing outside the browser canvas area.
2. Creation of Modal Windows.
3. File manipulation e.g. file creation, copying, renaming etc. this includes the possibility of formatting the hard disk, though we can't use that as evidence either.
4. Access to local data and operating system identifiers e.g. your username, your account login picture or whether or not you have accessed this website before.
5. Microsoft, User Account Control behaviour.
6. Existing best practice i.e. inspection of the TLS Certificate being used by your browser.

This is the original list with the exception of category 6 which was added after I had developed the solution. The quality of this list is irrelevant. I believed I could add to the list later, if necessary. Since the final solution is hidden within this list it was not necessary to add to it.

In my original research I dismissed or counterfeited every category except number 4. Every solution in Category 4 is actually a secret shared between the computer user and their web browser. With the exception of item number 6 this is the list from my original research. It has not been polished or edited. Item number 6 was added because this is current best practice. It is by accident that item number 4 just happened to contain the solution. Hence username, or account login picture, make good shared secrets, while previous access to this website is a bad shared secret. Previous access can be communicated via a darker colour hyperlink, or via browser dialogues such as the 'More Information' dialogue from the Firefox TLS window (Version 53.0.3). The darker colour hyperlink is easily counterfeited by any webpage. The browser dialogue can easily be counterfeited via full screen counterfeiting, documented below. Though the actual number of times you have accessed a website would be incorrect because Mallory does not have this information. It's still a bad signalling strategy because users don't track the number of times they have accessed a website.

Note, counterfeiting a browser dialogue with an undecorated window does not work anymore, see Figure 1. However, a floating DIV within a webpage can counterfeit a dialogue window i.e. on a webpage show a picture of a window, border and all. It's up to the user to notice that no window icon exists for this new window.

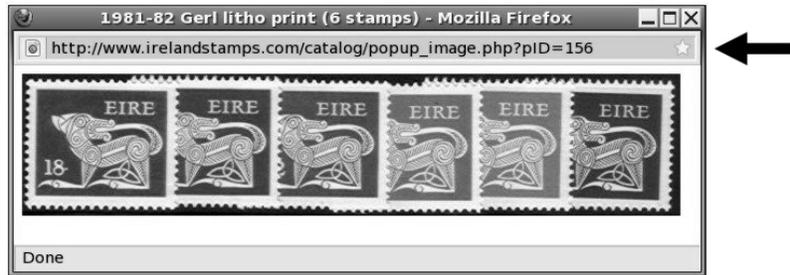

Figure 1. Attempt to create an undecorated window with JavaScript function 'window.open()'.

The arrow indicates an address bar added to prevent counterfeiting of browser controls.
Originally I dismissed Category 3 believing it to be unworkable. However references [3] and SiteKey [5] both use cookies to trigger their solutions. Cookies actually fit Category 3. These are 'cookie as a password' solutions. 'Cookie as a password' solutions fail because Alice-Human cannot successfully authenticate Bob, either at the regular login page or at the cookie creation page.

A key component of this research was the study of screening strategies. The actual path that I followed was to study the categories listed above. There is no point in me documenting that research here because it was straight forward and quite similar to discussions of screening strategies found in [1] and [2].

One phishing attack website that I stumbled upon requested a username and password. Even though the genuine website was open access. This type of phishing is more social engineering than counterfeiting. During my research I devised a versatile social engineering attack which allows the entire computer screen to be counterfeited, discussed next.

As stated, item number 6 was actually added after I had the solution. When I realised that even inspection of the TLS certificate could be counterfeited, in full screen mode.

## 2.1. Full Screen Counterfeiting

Full screen counterfeiting is easily achieved with a small amount of JavaScript and a set of bitmaps to fake the user's browser controls and desktop. Figure 2 shows six bitmaps set on a grey background. These images are deliberately drawn to appear fake, like crayon drawings.
Figure 4 shows a computer desktop, and browser, before a full screen counterfeiting attack. The 'Switch to Fullscreen!' button executes JavaScript. Figure 3 shows sample JavaScript code which implements the switch to full screen.

Description of Figure 2: Six bitmap pictures are shown on a grey background. The grey background is to help the reader see the size and shape of the bitmaps. The top three are to counterfeit browser controls while the bottom three show a counterfeit 'Windows start button', counterfeit taskbar with an application icon and clock. They are deliberately made to look fake, like crayon drawings. This is to help the reader see the difference between Figure 4 and Figure 6. The centre bitmaps will be tiled horizontally to help adjust the fake to any desktop resolution. The crayon like fake is made to look like the original NCSA Mosaic browser. An actual implementation would use 'navigator.userAgent' to ensure appropriate counterfeit images are presented.

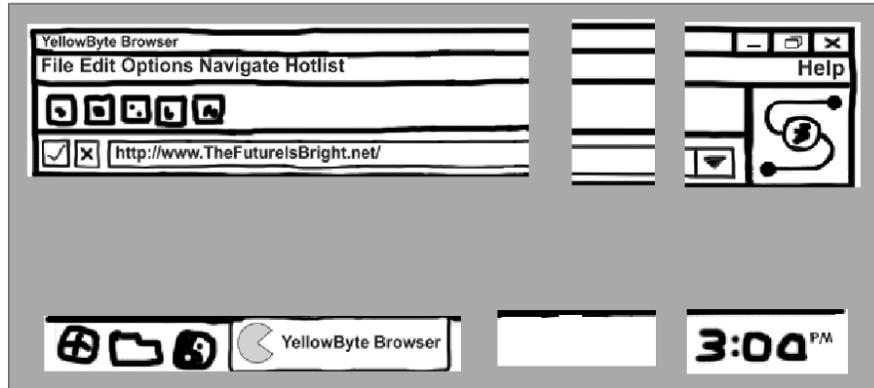

Figure 2. Six bitmaps used to create fake desktop. See text for description.

```
function Move_to_fullscreen() {
  //////////
  // Make the div visable....
  var qe = document.getElementById("crayon_browser");
  qe.style.visibility = "visible";
  qe.style.position   = "absolute";
  qe.style.left       = 0;
  qe.style.top        = 0;
  qe.style.width      = '100%';
  qe.style.height     = '100%';
  qe.style.zIndex     = "-1";

  //////////
  // Make the div fullscreen...
  var elem = document.getElementById("all_contents");
  if (elem.requestFullscreen) {
    elem.requestFullscreen();
  }

  // Another possible implementation, though commented out here, is...
  //
  //   if (elem.requestFullscreen) {
  //      elem.requestFullscreen();
  //   }
  //   else if (elem.mozRequestFullScreen) {
  //      elem.mozRequestFullScreen();
  //   }
  //   else if (elem.webkitRequestFullScreen) {
  //      elem.webkitRequestFullScreen();
  //   }
  //
  //
  //   document.onkeypress - requestFullscreen works for onkeypress.
  //   window.onload       - requestFullscreen does NOT work for
  //                         window.onload! Thankfully.
  //
...
```

Figure 3. Example of JavaScript code to switch browser to full screen mode.

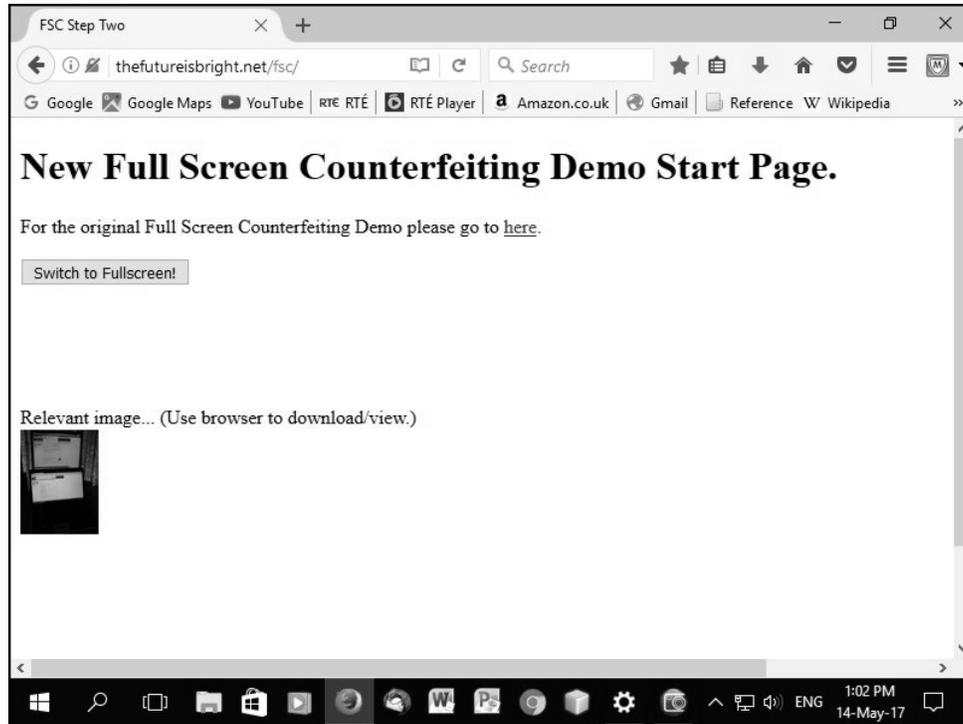

Figure 4. Screenshot of desktop before full screen counterfeiting attempt.

Basically a JavaScript function 'request Full screen ()' forces the browser into full screen mode. The same JavaScript code moves a HTML DIV to the front and makes it visible. That DIV 'crayon_browser' has the images from Figure 2 positioned in the corners or tiled to fit different desktop resolutions. It also contains the same webpage that was visible before the move to full screen.

Each web browser responds differently to the function 'request Full screen ()'. Figure 5 shows the warnings shown by three browsers. Microsoft Edge is both the best and worst. The warning shown in Figure 5 is shown the first time you switch to full screen. It stays on screen till the user dismisses it. This forces the user to explicitly acknowledge full screen mode. Unfortunately subsequent changes to full screen, on that website, do not warn the user at all i.e. Figure 4 is transformed directly to Figure 6 without any warnings. Firefox and Chrome show a warning every time. However these warnings dismiss themselves after a few seconds. Aside from the different transition warnings, shown in Figure 5, all three browsers transform Figure 4 into Figure 6.

If the bitmaps used in Figure 2 were realistic then Figure 4 and Figure 6 would be almost identical. Furthermore the transition warnings shown in Figure 5 would only appear odd/unusual because they appeared outside of the perceived canvas area. These are very weak indicators of counterfeiting.

From a researcher's point of view many types of installed software behaviour can be counterfeited. Including browser addons, inspection of TLS certificates, and Microsoft User Account Control behaviour. As such categories 5 and 6 must be eliminated as suitable anti-counterfeiting solutions. Furthermore we now need to be concerned with counterfeiting of blue screens of death, hackers/criminals blackmailing people with the threat of formatting their hard drives etc.

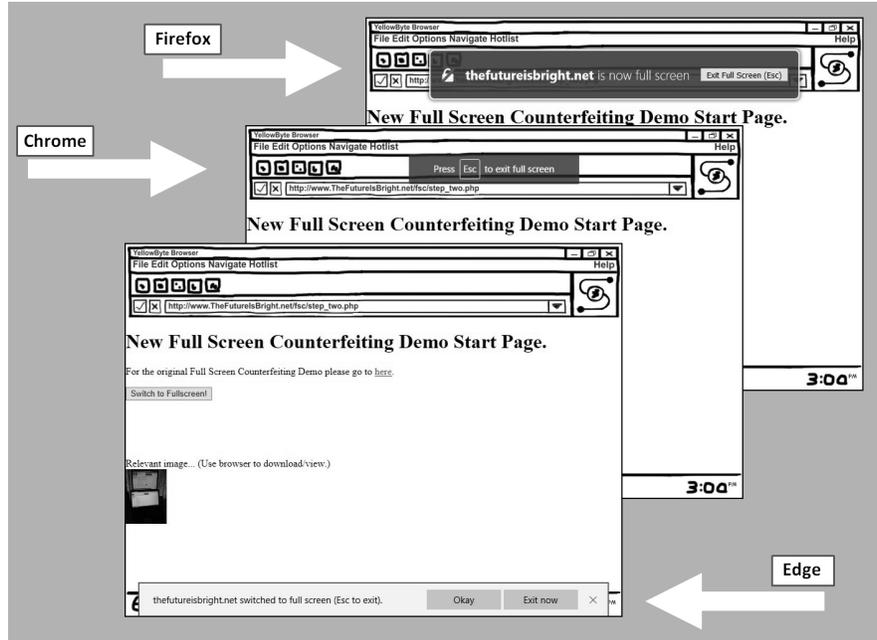

Figure 5.  Shown are the warnings presented by three browsers after the JavaScript function 'requestFullscreen()' is called.

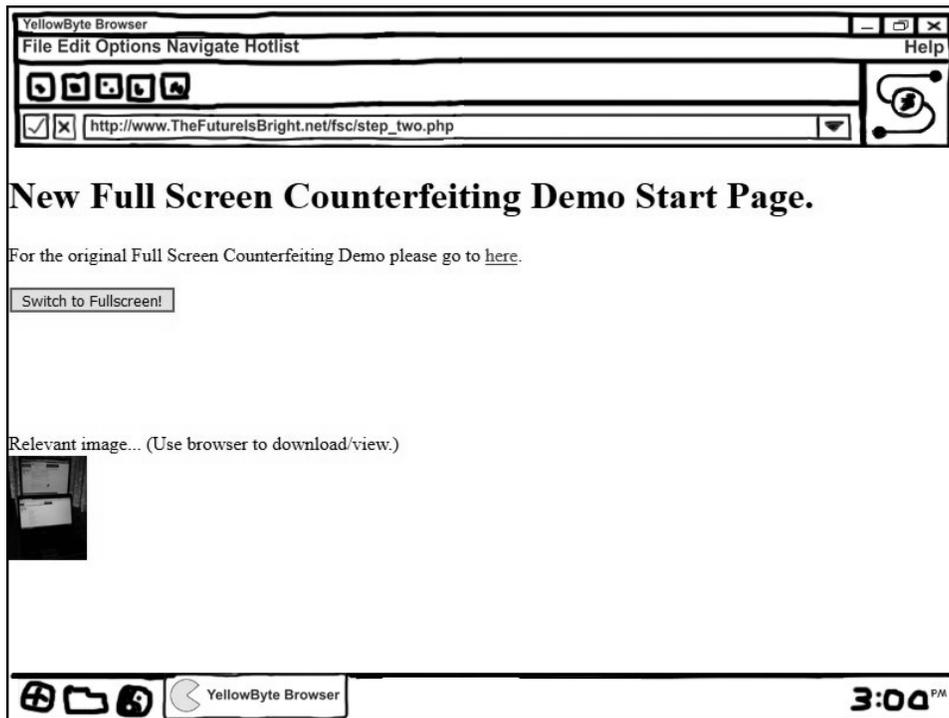

Figure 6.  Screenshot after full screen counterfeiting attack, compare with Figure 4. Also see the bitmaps in Figure 2, used to counterfeit the desktop.

The purpose here is to demonstrate these mechanisms. No user testing has been performed. There is anecdotal evidence in [6] that these tactics will work. The academic exercise of demonstrating that this is possible is sufficient to eliminate categories 5 and 6. What is of interest is the inability of this mechanism to counterfeit category 4 solutions. It is this fact which suggests the hypothesis proposed in this paper.

## 3. PROPOSED HYPOTHESIS

Figure 7 shows a login dialogue which embodies the solution. The only behaviour which cannot be counterfeited by full screen counterfeiting is the presentation of previously shared secrets i.e. a cryptography authentication mechanism since the time of antiquity. The sections which follow document various aspects of the proposed hypothesis.

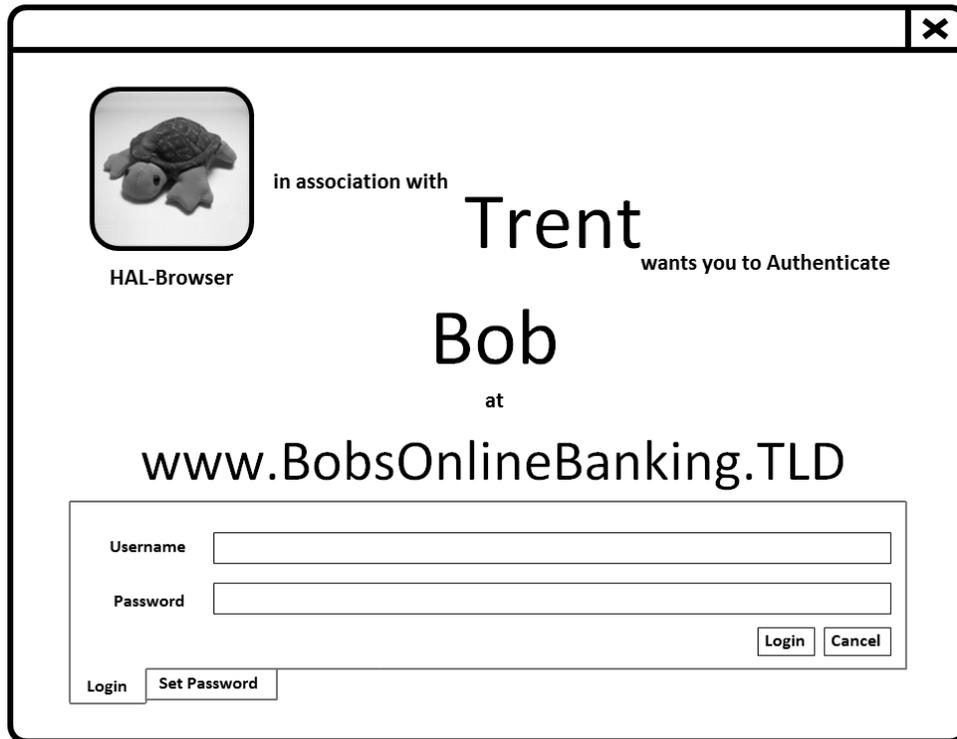

Figure 7.  Window for user to authenticate (1) their own browser, via picture (2) Bob, (3) Bob's website, (4) Trent, via browser confirmation of the digital signature and finally login fields. The use of which indicates acceptance of the various credentials presented.

Description of Figure 7: This is not SiteKey. This is not a webpage. This is a browser created dialogue. Created with a user-browser shared secret, obtained from the hard disk, and identity credentials from the TLS certificate. For Mallory to carry out a MITM attack she must stand between you and your computer monitor. Either that or hack into your computer to steal the shared secret. Hacking into thousands of computers to steal shared secrets is an entirely different endeavour to creating a counterfeit website. Furthermore having hacked into your computer why bother with a phishing attack?

In a nutshell: Your browser is a virtual machine. Each webpage executes inside its own sandbox. Once your view the installed software component, of your browser, as an actor in the

23

cryptography protocol, everything else is classic cryptography i.e. your browser seeks authentication of Bob from Alice-Human. This act is vulnerable to counterfeiting. As such your browser utilises a signalling strategy to communicate that it is the correct actor i.e. it reads a shared secret from the hard disk and presents it to Alice-Human. Sandboxed processes, websites, cannot do this. Hence Alice-Human can interpret the correct shared secret as proof that the browser created the window, rather than a sandboxed website a.k.a. a phishing website.

Figure 7 should be displayed as a modal window, positioned in the middle of the screen. If the rest of the screen can be greyed, like Microsoft User Account Control, then even better. Arguably in Figure 7 putting the login fields into a dialogue with the browser signals, TLS identity, is more important than the shared secret i.e. it forces Alice-Human to look at the Padlock, or green bar from extended validation TLS. I just happened to use the TLS identity rather than a padlock, or green bar. Furthermore, the shared secret prevents the phishers from making their next move i.e. to counterfeit Figure 7.

## 4. TWO ACTORS OR THREE?

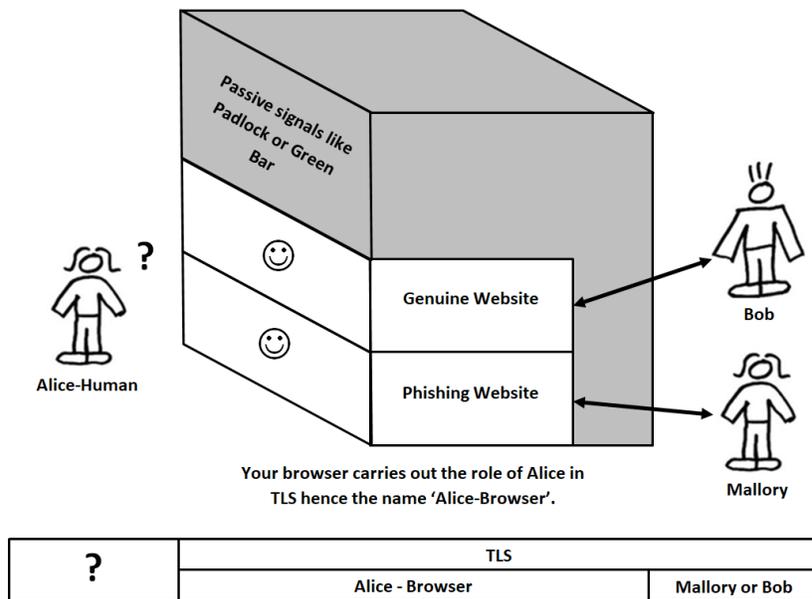

Figure 8. Two actors: The browser is a virtual machine with each website sandboxed inside it. The figure represents the current situation where TLS is implemented by two actors i.e. Alice-Browser and either Bob or Mallory. Alice-Human plays a passive role. By default she accepts the TLS identity without being forced to examine it. She must remember to look for the passive signals from the browser.

In the current system Alice-Human's participation in TLS is optional. The sandboxed websites look the same. See Figure 8. The signals from the browser are passive and displayed away from the main event, the webpage. In Figure 9 the response to successful authentication of the TLS certificate is not to display a padlock, or a green bar, rather to display Figure 7. The green bar can be ignored by Alice-Human. Figure 7 cannot be ignored. Alice-Human must enter her login credentials into it. Or, into a counterfeit of it. It's an active process rather than passive. Part of this solution is that regular webpages will no longer provide login fields i.e. you will only enter your username and password into the dialogue in Figure 7. Request of login credentials, on a regular webpage, should be viewed as suspicious by users.

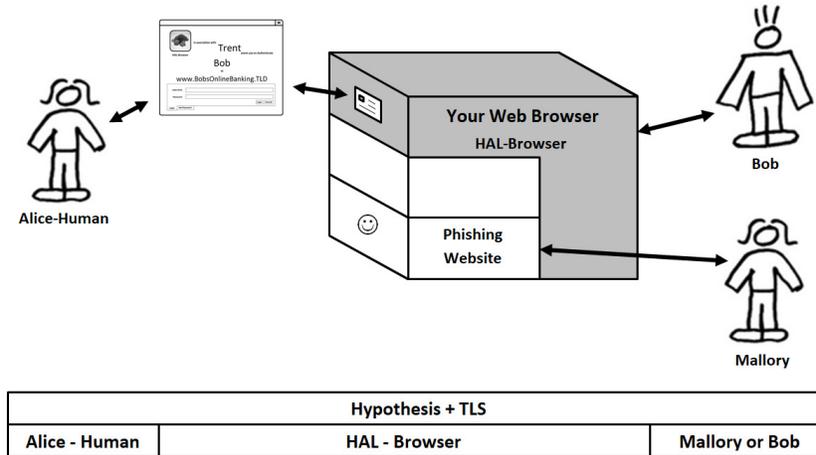

| Hypothesis + TLS | | |
|---|---|---|
| Alice - Human | HAL - Browser | Mallory or Bob |

Figure 9. Three actors: The virtual machine itself must participate in TLS, not a sandboxed process. The current implementation of TLS shows a padlock, or green bar, on successful verification of a TLS certificate. With this solution the browser shows Figure 7 on successful verification of a TLS certificate. Only after Alice-Human enters her login credentials will the browser proceed to create the website inside its own sandbox. Your browser can access the hard disk, so it can place the correct shared secret on the dialogue. Sandboxed websites cannot access the hard disk. Hence they cannot counterfeit Figure 7.

In the existing system, Figure 8, your browser fulfils the role of Alice. In the proposed solution Alice is the human being sitting at the computer. To aid discussion I have used the names Alice-Human, Alice-Browser and HAL-Browser. Alice-Browser refers to the current situation where your browser fulfils the role of Alice within TLS. The human is present but her role is, at best, passive. In the proposed solution Alice-Human plays an active role, authenticating both her browser window and the TLS identity.

In the existing system Alice-Browser verifies the digital signature on Bob's TLS certificate. On success Alice-Browser and Bob proceed to implement TLS i.e. two actors. In the new model, Figure 9, HAL-Browser verifies the digital signature on Bob's TLS certificate. On success HAL-Browser turns to Alice-Human and invites her to further authenticate Bob. He does this by displaying Figure 7. The problem is: this act is vulnerable to counterfeiting. In this context counterfeiting is referred to as a phishing attack.

Shown in Figure 7 is a picture of a turtle which is a shared secret between Alice-Human and HAL-Browser. Neither Bob nor Mallory know this secret. As such Mallory cannot counterfeit Figure 7 without hacking into HAL-Browser to steal the secret. Hacking into thousands of computers to steal these secrets is an entirely different endeavour to tricking people into going to a fake website.

Once you correctly model the system as a three actor system. Cryptographers know how to appropriately authenticate the three participants. As such Figure 7 is a relatively obvious step for cryptographers. Dhamija et al also use a user-display shared secret. They use it to protect a dedicated login window from counterfeiting. They do not appear to go beyond that and use it to present Bob's identity credentials [4]. With my solution, by entering her login credentials Alice-Human is accepting Bob's identity credentials and her browser's shared secret. She is authenticating both Bob and her web browser. HAL-Browser then proceeds to implement TLS.

25

Hence Figure 7 extends TLS by forcing Alice-Human to carry out these additional authentication steps.
Alice-Human now knows she is looking at a dialogue created by her web browser i.e. it is not a counterfeit, a phishing attack. She can now examine the identity credentials presented and complete Bob's authentication.

TLS would need to be modified to implement the solution e.g. websites should be able to choose 'no login dialogue', 'no set password tab', among other possibilities. I was approaching this as a game theorist seeking screening strategies to prevent counterfeiting. Here follows an outline of the game theory interpretation.

### 4.1. Shared Secret Authentication as a Screening Strategy

Anti-counterfeiting technologies and the screening strategy that accompany them go together like a lock and key pair. The research involved the study of each category, from section 2, to find screening strategies which would prevent phishing attacks.

The definition of a screening strategy, from [2] is given since its language is used to frame the discussion that follows. From [2]: 'A screening strategy is a strategy used by a less informed player to elicit information from a more informed player'.

Human Interactive Proofs (e.g. CAPTCHA), Turing tests and anti-counterfeiting technologies are all specific types of screening strategy. Here too authentication, through the confirmation of a shared secret, constitutes a screening strategy. The less informed player is eliciting the identity of the more informed player. They are not eliciting the secret because they already know it. They want to know 'do you know what the secret is?' This is why it's just a point of view that this is cryptography. As a game theorist I see a screening strategy. It elicits their identity, as the individual who knows the secret or someone else.

Furthermore, the fact that this works while other approaches fail indicates phishing attacks involve the counterfeiting of an identity, not a website. This is significant because it allows us to prevent any type of counterfeiting. It recasts counterfeiting as theft of intellectual property, patents, copyright, trademarks, designs etc. accompanied by identity theft. The purpose of the identity theft is to undermine law enforcement attempts which would otherwise prevent the intellectual property theft. This means authentication based solutions can be developed for any type of counterfeiting including manufactured goods like pharmaceutical drugs and currencies.

## 5. ADDITIONAL SOLUTION DETAILS

The proposed solution is to display Figure 7 on successful verification of a TLS certificate's digital signature. The key points are:

1. It's the installed software component of your browser which does this. Not a sandboxed website. Nor is this a webpage hosted somewhere on the internet. That would be SiteKey. This is not SiteKey.
2. Alice-Human elicits the identity of whoever created Figure 7 through a screening strategy i.e. sandboxed websites cannot access the hard disk whereas the virtual machine, your browser, can.
3. With the current situation it's up to Alice-Human to remember to check for a padlock and/or green bar. The default is for Alice-Human to accept or reject Bob/Mallory based upon the website's contents. With the proposed solution Alice-Human cannot ignore the two identities being presented in Figure 7. She must examine Figure 7 in order to enter

   her login credentials. She does not have to remember to check these identities nor is the default, automatic acceptance, when she forgets to check for a padlock symbol.
4. Microsoft user account control behaviour can be used to further enhance the solution. If Figure 7 is displayed as a modal dialogue with the rest of the screen blanked. This will undermine even more attacks e.g. to counterfeit a modal window a phishing website need only create an image of a dialogue window. Then position that image on their phishing webpage as if it's a real dialogue. Real windows would create a window icon, in the operating system. Users who don't notice the absence of a window icon may be tricked into using that dialogue. Such fake login screens would have an incorrect authentication image (the turtle in Figure 7).
5. Where Mallory buys/obtains a TLS certificate Figure 7 will be displayed with the correct authentication image and whatever data is stored inside the certificate. If this solution is adopted then a large number of issues with TLS certificates and certificate authorities will need to be resolved.
6. In Figure 9 the genuine website is absent. This is because the installed software component of your browser will only create the sandboxed website after a secure TLS connection has been created. Hence Figure 9 shows the point just before Alice-Human has entered her login credentials and clicked 'Login'.
7. Central Banks as Trent: When users are looking at Figure 7 it will become apparent that the public have never heard of any of the Certificate Authority companies. And who will trust a Trent they have never heard of? One solution is for central banks to fulfil the role of Trent within their regulatory area. Hence the Federal Reserve, the European Central Bank etc. should fulfil the role of Trent. The actual task of creating TLS certificates can be outsourced to a Certificate Authority. The name for Trent in Figure 7 should be a name the public know and trust.
8. While a patent application was filed [7] this application has now lapsed. Specifically all patent deadlines have now lapsed including USA, Canada etc. where applications can be made up to one year after publication of an idea. The solution is now prior art everywhere in the world.

## 6. RELEVANT METAPHORS AND ANALOGIES

### 2.2. April fool's day at a TV Station

Consider the following: its April fool's day and someone in a television station decides to play a joke on their viewers. They pick a popular brand of television, counterfeit it's setup menu and then superimpose that image over the live television broadcast. Viewers who own a different brand of television will be like a Bank of Ireland customer receiving a Bank of America phishing email i.e. they will know immediately that it's a scam. However, viewers with the correct brand of television will think their television is malfunctioning as it is presenting the setup menu no matter what they do. To prevent this trick from working, viewers must customise their setup menu. Doing so is creating a secret known by their television and themselves, but not known by the television station. This is identical to the solution to phishing attacks i.e. Mallory cannot counterfeit what Mallory does not know. It's a viewer-television secret just like our browser-user secret.

### 2.2. HAL, friend or foe?

In 2001, a Space Odyssey HAL had two personalities, one friend one foe. Imagine that we give the friend personality an Identity Card which he should present when we're talking to him, to help us distinguish friend from foe. Effectively that is the solution presented i.e. when the installed software is acting on our behalf it has access to the shared secret. When a remote website is counterfeiting a website it cannot present a fake TLS certificate nor can it fake the shared secret.

Computer users must authenticate both their web browser and the identity presented in the TLS certificate. This is where the name HAL-Browser came from. Our web browsers have split personalities one friend one foe. The user-browser shared secret is an identity card for our friend.

## 7. CONCLUSIONS

Once the installed software component, of your browser, is recognised as an actor in the cryptography protocol everything else is classic cryptography i.e. it must authenticate itself by presenting a previously shared secret. Otherwise a sandboxed website will counterfeit it i.e. a phishing attack. Sandboxed websites cannot access the hard disk, hence they cannot counterfeit Figure 7. After that your browser's participation in the TLS protocol is textbook three actor interaction. On successful authentication of a TLS certificate's digital signature. HAL-Browser seeks further authentication from Alice-Human. This step involves HAL-Browser authenticating himself with Alice-Human through the presentation of a previously shared secret. This step also involves HAL-Browser presenting Bob's identity credentials from the TLS certificate. Alice-Human can accept these two identities and enter her login credentials or she can reject either of the identities presented and back away, refusing to enter her login credentials.